\documentclass[amsmath,amssymb]{revtex4-1}
\usepackage{graphicx}
\usepackage{bm}
\usepackage{amsthm} 
\usepackage{natbib} 

\begin{document} 
\title[Survival of interacting Brownian particles]{Survival of interacting Brownian particles in crowded 1D environment}
\author{Artem Ryabov}
\email{rjabov.a@gmail.com}
\author{Petr Chvosta}
\affiliation{Department of Macromolecular Physics, Faculty of Mathematics and Physics, Charles University in Prague, V Hole{\v s}ovi{\v c}k{\' a}ch 2, 180~00~Praha, Czech Republic}
\date{\today}
\begin{abstract}
We investigate a diffusive motion of a system of interacting Brownian particles in quasi-one-dimensional micropores. In particular, we consider a semi-infinite 1D geometry with a partially absorbing boundary and the hard-core inter-particle interaction. Due to the absorbing boundary the number of particles in the pore gradually decreases.
We present the exact analytical solution of the problem. Our procedure merely requires the knowledge of the corresponding single-particle problem.
First, we calculate the simultaneous probability density of having still a definite number $N-k$ of surviving particles at definite coordinates. Focusing on an arbitrary tagged particle, we derive the exact probability density of its coordinate.
Secondly, we present a complete probabilistic description of the emerging escape process. The survival probabilities for the individual particles are calculated, the first and the second moments of the exit times are discussed.
Generally speaking, although the original inter-particle interaction possesses a point-like character, it induces entropic repulsive forces which, e.g., push the leftmost (rightmost) particle towards (opposite) the absorbing boundary thereby accelerating (decelerating) its escape. More importantly, as compared to the reference problem for the non-interacting particles, the interaction changes the dynamical exponents which characterize the long-time asymptotic dynamics.
Interesting new insights emerge after we interpret our model in terms of a) diffusion of a single particle in a $N$-dimensional space, and b) order statistics defined on a system of $N$ independent, identically distributed random variables.
\end{abstract}
\maketitle
\section{\label{sec:intro}Introduction}

Diffusion dynamics of hard-core interacting particles in one dimension (1D) is known as the single-file diffusion (SFD).
The SFD occurs for instance in narrow biological pores\cite{Hodgkin}, inside the zeolites\cite{HahnKarger,KargerBOOK}, during the sliding of proteins along the DNA\cite{DNAnature, Hippel}. Other examples comprise migration of adsorbed molecules on surfaces\cite{Halpin}, diffusion in nanotubes\cite{nanoWaterI, nanoWaterII}, carrier migration in polymers and superionic conductors\cite{Superionic, SuperionicII}. Advances in technology now allow to study artificial single-file systems such as solid-state nanopores\cite{SSnanopores}, confined colloid particles\cite{BechCircle, BechPhoto, colloidpolym}, charged spheres in circular channel\cite{ChargedSpheres}, or millimetric steel balls in plane condenser\cite{BallsOFsteel}. In all these systems, the mutual passage of the particles is forbidden which results in deviation of tagged particle's dynamics from the classical diffusion.

The concept of SFD was first introduced in the mathematical literature by Harris\cite{Harris}. He proved that the mean-square displacement of the tagged particle increases with time as $t^{1/2}$ in contrast to its linear increase for the free particle. For a comprehensible review of the present state of the art in the theory of SFD cf.\ Introduction in Ref.\ \onlinecite{BarSil2010}). Despite the long history of SFD phenomena, exactly solvable models of single-file systems are rare and they have occurred only recently.
The exact solution of the diffusion equation for $N$ hard-core interacting particles has been first derived in Ref.\ \onlinecite{KargerHahn} for the diffusion on an infinite line
(it has been rederived in Ref.\ \onlinecite{Kumar}). The SFD of $N$ particles within the finite interval has been solved through the Bethe ansatz in Ref.\ \onlinecite{SFDLizana}.
In Ref.\ \onlinecite{RC2011}, the present authors have generalized the previously known exact solutions to the case of an arbitrary time- and space-dependent force acting on particles. All these exact solutions can describe real physical situations, however, they all concern the systems with a conserved number of particles.
Often the opposite is true and the single-file systems should be treated as open systems with \emph{variable} number of particles.
Many such systems are listed in the first paragraph. The single-file conditions in combination with a variable number of particles are two prerequisites for a proper understanding of kinetics of diffusion-limited catalytic reaction in crowded environments\cite{DaJiangLiu}. Also the search for corrections to Smoluchowski theory which would incorporate the effects of excluded volume\cite{Smoluch, JoonhoPark} stimulates the interest in one-dimensional models\cite{Kazuhiko} with hard-core interaction and particle absorption.

To our knowledge, the only work that presents analytical study of dynamics of a tagged particle in an open system is Ref.\ \onlinecite{KargerHahn} where the finite interval with two perfectly absorbing boundaries was briefly discussed. However, the approximative tagged particle probability density function (PDF) obtained in the thermodynamic limit was only considered. In the present paper we develop a new general formalism that allows an exact analytical treatment of the SFD problem with the absorbing boundary (such boundary can represent, e.g., the open end of the biological pore, a fixed reactant, a binding site inside the pore, etc.).
The approach is a logical outgrowth of our earlier method\cite{RC2011} and its characteristic feature is that the exact solution of the SFD problem can be given whenever
the exact solution of the corresponding single-particle problem is known. Formally, our analysis is based on a hierarchical system of coupled generalized diffusion equations describing the evolution periods when the pore contains a definite number of particles. We were able to solve the hierarchy. The solution yields the complete information concerning the many-particle dynamics. The physical consequences of the particle interaction manifest themselves most transparently if we follow the motion of an individual tagged particle. We were able to derive the exact expression for the tagged particle PDF and for its survival probability. The hard-core inter-particle interaction induces entropic forces
which radically alter the character of the escape process comparing to the corresponding problem of noninteracting particles.

The paper is organized as follows. Sec.\ \ref{sec:model} contains the definition of the model. Sec.\ \ref{sec:evolutioneqs} comprises four subsections. First, for the sake of a latter reference, Subsec.\ \ref{subsec:singleparticle} presents the complete solution of the underlying single-particle case. In Subsec.\ \ref{subsec:evolutioneqs} we formulate the hierarchy of coupled generalized diffusion equations for the partial PDFs. The appropriate choice of the initial conditions is discussed in Subsec.\ \ref{subsec:initialconditions}. In Subsec.\ \ref{subsec:mapping} the SFD dynamics is translated into the diffusion of a single compound particle in a special $N$-dimensional domain.
Sec.\ \ref{sec:solution} contains the exact solution of our model (some purely technical steps are particularized in appendixes). The Eqs.\ (\ref{theorem1}) and (\ref{theorem2}) in Subsec.\ \ref{subsec:partialPDFs} give the exact solution of the hierarchy of coupled diffusion equations from which the tagged particle's PDF (Eq.\ (\ref{marginalf})) is derived in Subsec.\ \ref{subsec:taggedPDFs}. We then turn, in Sec.\ \ref{sec:escapeprocess}, to a detailed analysis of the escape process.
Subsec.\ \ref{subsec:survival} discusses the survival probabilities and their asymptotic behavior. The exit times are analyzed in Subsec.\ \ref{subsec:escapetime}.
\section{\label{sec:model}Definition of the model}
\begin{figure}
\includegraphics[scale=1]{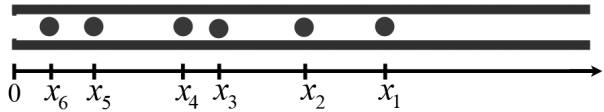}
\caption{An illustration of the one-dimensional semi-infinite pore containing $N=6$ diffusing particles and the partially absorbing boundary at the origin (the exit from the pore). The diffusing particles interact as hard spheres and hence their initial ordering $x_{6}<x_{5}< ... < x_{2}<x_{1}$ is retained during the time evolution.}
\label{fig:singlefile}
\end{figure}

Consider $N$ identical Brownian particles, each with the diffusion constant $D$, diffusing in the one-dimensional semi-infinite pore. The particles are subject to the following two interactions:
\begin{itemize}
\item[({\it i})] \emph{The particle-particle hard-core interaction.}
As the consequence, particles cannot pass each other. Their initial ordering $x_{N}<x_{N-1}< ... <x_{1}$ is preserved during the subsequent diffusion process.
\item[({\it ii})] \emph{The particle-boundary interaction.}
We assume a partially absorbing boundary at $x = 0$. The particle that hits the boundary is either removed from the pore, or it is immediately reflected and it continues to diffuse inside the pore. The boundary will be implemented through the radiation boundary condition \citep{Redner}. The ``particle-capture capacity'' of the boundary is characterized by the parameter $\kappa$. The value $\kappa=0$ corresponds to a reflecting boundary, the limit $\kappa \to \infty$ yields a totally absorbing boundary.
\end{itemize}
The particles cannot enter the pore from the outside and they can leave it only through the boundary at the origin. The number of particles inside the pore either decreases or remains constant over time.
\section{\label{sec:evolutioneqs} Evolution equations}
\subsection{\label{subsec:singleparticle} Single-particle case}

In order to introduce basic concepts needed in the following, let us first consider the simplest situation $N=1$. That is, at the initial time $t=0$, there is only \emph{one} particle  in the pore and it is located at the position $y$, $y>0$. The initial PDF $p(x;0\,|\,y;0)$ for the particle position is the Dirac delta function
\begin{equation}
\label{initialcond}
p(x;0\,|\,y;0) = \delta(x-y)\,\,.
\end{equation}
Inside the pore (i.e., for $x >0$), the time-evolution of the initial condition (\ref{initialcond}) is governed by the diffusion equation
\begin{equation}
\label{diff0}
\frac{\partial}{\partial t} p(x;t\,|\,y;0)  = D \frac{\partial^{2} }{\partial x^{2}} p(x;t\,|\,y;0)\,\,.
\end{equation}
At the boundary, we require the radiation boundary condition (BC)
\begin{equation}
\label{boundarycond}
\left.\left(D\frac{\partial}{\partial x}-\kappa\right)p(x;t\,|\,y;0)\right|_{x=0} = 0\,\,.
\end{equation}
The exact solution of Eq.\ (\ref{diff0}) with the initial condition\ (\ref{initialcond}) and BC\ (\ref{boundarycond}) reads
\begin{eqnarray}
\notag
p(x;t\,|\,y;0) &=&  \frac{\exp{\!\! \left[\!- (x-y)^{2}/(4Dt) \right]}}{\sqrt{4 \pi D t}} +
\label{Psol}
\frac{\exp{\!\!\left[\!- (x+y)^{2}/(4Dt)\right]}}{\sqrt{4 \pi D t}}-\\
\notag
&&- \frac{\kappa}{D}\exp{\!\! \left[\!\frac{\kappa}{D}(x+y)\!+\!\kappa^{2}\!\frac{t}{D}\right]}
{\rm{erfc}}\!\! \left(\frac{x+y}{\sqrt{4Dt}}+\kappa\sqrt{\frac{t}{D}}\right),
\end{eqnarray}
where $\rm{erfc}(\bullet)$ stands for the complementary error function \citep{AbrStegun}. It can be derived, e.g., by the Laplace transform method\citep{Duffy}.

Let $f(x)$, $x>0$, be an arbitrary PDF describing the initial particle position. Having calculated the Green function (\ref{Psol}), the time evolution of the initial condition
is given as
\begin{equation}
\label{pDEF}
p(x;t)=\int_{0}^{+\infty}\!\!\!\! {\rm d}y\, p(x;t\,|\,y;0) f(y)\,\,.
\end{equation}
The probability of finding the particle at the time $t$ within the interval $(0,x)$ is given by the cumulative distribution function
\begin{equation}
\label{FDEF}
F(x,t)=\int_{0}^{x}\! {\rm d}x' p(x';t)\,\,.
\end{equation}
The probability that the particle has not exited the pore by the time $t$, i.e., its survival probability, reads
\begin{equation}
\label{survprob}
S(t)=\lim_{x \rightarrow + \infty} F(x,t) \,\,.
\end{equation}
It is an important feature of our setting that all subsequent $N$-particle results can be formulated through the single-particle functions $p(x;t)$, $F(x,t)$, $S(t)$, and only through them.
\subsection{\label{subsec:evolutioneqs} $N$-particle case}

Suppose that, at the initial time $t = 0$, there are $N$ particles in the pore and that their positions
$0<x_{N} < x_{N-1} <\, ...< x_{1}$, cf.\ Fig.\ 1, are distributed according to the joint PDF $f^{(N)}(\vec{x}_{N};0)$, $\vec{x}_{N}=(x_{1},...\,,x_{N})$.
Having in mind that the total number of particles inside the pore \emph{is not conserved}, we introduce partial $(N-k)$-particle PDFs $f^{(N-k)}(\vec{x}_{N-k};t)$, $k=1, ...\, , N-1$,
$\vec{x}_{N-k}=(x_{1},...\,,x_{N-k})$. We give them the following meaning:
$f^{(N-k)}(\vec{x}_{N-k};t){\rm d}\vec{x}_{N-k}$ is the probability that, at the time $t$, there are $N-k$ particles in the pore (this is stressed by the superscript)
\emph{and} that these particles are located in $(\vec{x}_{N-k},\vec{x}_{N-k}+{\rm d}\vec{x}_{N-k})$. Altogether, we have $N$ partial PDFs. Further, let $f^{(0)}(t)$ be
the probability that, at the time $t$, there are no particles left in the pore. The knowledge of the partial PDFs and of the probability $f^{(0)}(t)$ represents the complete probabilistic description of the resulting single-file diffusion problem.

The partial PDF $f^{(N)}(\vec{x}_{N};t)$ varies in time due to the following two reasons: a) the motion of $N$ interacting particles, b) the possible exit of the particle No.\ $N$
out of the pore. The diffusion equation
\begin{equation}
\label{fnEQ}
\frac{\partial}{\partial t} f^{(N)}(\vec{x}_{N};t) = D\sum_{j=1}^{N} \frac{\partial^{2} }{\partial x_{j}^{2}} f^{(N)}(\vec{x}_{N};t)\,\,,
\end{equation}
together with the ``non-crossing'' boundary conditions
\begin{equation}
\left(\frac{\partial}{\partial x_{j}} -  \frac{\partial}{\partial x_{j+1}} \right)
\left. f^{(N)}(\vec{x}_{N};t)\right|_{x_{j}=x_{j+1}} = 0\,\,,
\label{nNC}
\end{equation}
$j=1,...\,N-1$, account for the diffusion of $N$ hard-core interacting particles inside the pore. The non-crossing boundary conditions (\ref{nNC}) can be derived by introducing the $N$-particle probability {\em current\/} and by assuming that the normal component of the current vanishes at the hyperplanes $\left\{\vec{x}_{N}:0 \leq x_{N} < ...<x_{n}=x_{n-1}<...<x_{1}< + \infty  \right\}$, $n=2,...,N$. Their derivation and the further discussion can be found in Ref.\ \onlinecite{RC2011}, Eqs.\ (3)-(5). BCs (\ref{nNC}) ensure that the initial ordering of the particles is preserved. Hence, it is only the leftmost particle (the particle No.\ $N$) that can leave the pore.
The exit process is incorporated by the radiation BC (cf.\ Eq.~(\ref{boundarycond}))
\begin{equation}
\left(D\frac{\partial\phantom{0}}{\partial x_{N}}  - \kappa \right) \left. f^{(N)}(\vec{x}_{N};t)\right|_{x_{N}=0} = 0\,\,.
\label{nRC}
\end{equation}
The diffusion equation (\ref{fnEQ}) must be supplemented by an initial condition $f^{(N)}(\vec{x}_{N};0)$. This will be done in Subsec.\ \ref{subsec:initialconditions}.

Let us now focus on the partial PDF $f^{(N-1)}(\vec{x}_{N-1};t)$. It varies due to the following \emph{three} reasons:
a) the motion of $N\!-\!1$ interacting particles, b) a possible exit of the particle No.\ $N-1$ out of the pore, c) the possible exit of the particle No.\ $N$ while the remaining particles are situated in the infinitesimal domain $(\vec{x}_{N-1},\vec{x}_{N-1}+{\rm d}\vec{x}_{N-1})$. We thus arrive at an \emph{inhomogeneous} generalized diffusion equation
with a source term on its right-hand side:
\begin{eqnarray}
\frac{\partial}{\partial t} f^{(N-1)}(\vec{x}_{N-1};t) =
D  \sum_{j=1}^{N-1} \frac{\partial^{2} }{\partial x_{j}^{2}} f^{(N-1)}(\vec{x}_{N-1};t) 
  + \kappa \left. f^{(N)}(\vec{x}_{N};t)\right|_{x_{N}=0}.
\label{fn-1EQ}
\end{eqnarray}
The last term on the right-hand side accounts for the above point c). The hard-core interaction of $N-1$ particles is again incorporated through non-crossing BCs
\begin{equation}
\left(\frac{\partial}{\partial x_{j}} -  \frac{\partial}{\partial x_{j+1}} \right)
\left. f^{(N-1)}(\vec{x}_{N-1};t)\right|_{x_{j}=x_{j+1}} = 0\,\,,
\label{n-1NC}
\end{equation}
$j=1,...\,N-2$. The radiation BC describing the possible escape of the particle No. $N\!-\!1$ reads
\begin{equation}
\left(D\frac{\partial\phantom{0}}{\partial x_{N-1}}  - \kappa \right) \left. f^{(N-1)}(\vec{x}_{N-1};t)\right|_{x_{N-1}=0} = 0\,\,.
\label{n-1RC}
\end{equation}
Presently, the initial condition supplementing Eq.\ (\ref{fn-1EQ}) is simply $f^{(N-1)}(\vec{x}_{N-1};0)=0$. This follows from the probabilistic interpretation of PDF $f^{(N-1)}(\vec{x}_{N-1};t)$ and from the assumption that, at $t=0$, there are $N$ particles inside the pore.

Following a similar reasoning, the generalized diffusion equation that governs the evolution of the partial PDF $f^{(N-k)}(\vec{x}_{N-k};t)$, $k=2,3,\ldots,N-1$, reads
\begin{eqnarray}
\frac{\partial}{\partial t} f^{(N-k)}(\vec{x}_{N-k};t)  =
D  \sum_{j=1}^{N-k} \frac{\partial^{2} }{\partial x_{j}^{2}} f^{(N-k)}(\vec{x}_{N-k};t) +
 \kappa  \left. f^{(N-k+1)}(\vec{x}_{N-k+1};t)\right|_{x_{N-k+1}=0}.
\label{fn-kEQ}
\end{eqnarray}
Again, the equation is supplemented by the BCs
\begin{equation}
\label{n-kNC}
\left(\frac{\partial}{\partial x_{j}} -  \frac{\partial}{\partial x_{j+1}} \right)
\left. f^{(N-k)}(\vec{x}_{N-k};t)\right|_{x_{j}=x_{j+1}} = 0\,\,,
\end{equation}
$j=1,...\,N-k-1$, by the radiation BC
\begin{equation}
\left(D\frac{\partial\phantom{0}}{\partial x_{N-k}}  - \kappa \right) \left. f^{(N-k)}(\vec{x}_{N-k};t)\right|_{x_{N-k}=0} = 0\,\,,
\label{n-kRC}
\end{equation}
and by the initial condition $f^{(N-k)}(\vec{x}_{N-k};0)=0$. Of course, in the case $k=N-1$, there is no non-crossing BC.

In the last step of the recursive scheme we focus on the time evolution of the probability $f^{(0)}(t)$. It increases due to the escape of the last remaining particle and it obeys the simple evolution equation
\begin{equation}
\label{f0EQ}
\frac{\rm d}{{\rm d} t} f^{(0)}(t) = \kappa \left. f^{(1)}(x_{1};t)\right|_{x_{1}=0},
\end{equation}
with the initial condition $f^{(0)}(0)=0$.

Summing up the above steps, we have obtained a \emph{hierarchy of $N+1$ coupled evolution equations} (\ref{fnEQ}), (\ref{fn-kEQ}) for $k=1, ..., \, N-1$, and (\ref{f0EQ}).
The diffusion equation (\ref{fnEQ}) is \emph{closed}. Together with Eqs.\ (\ref{nNC}), (\ref{nRC}) and with the initial condition $f^{(N)}(\vec{x}_{N};0)$ it constitutes
a well-defined initial-boundary value problem with a unique solution, the partial PDF $f^{(N)}(\vec{x}_{N};t)$. Suppose that we know it.
Then we insert its value at $\left. \vec{x}_{N}\right|_{x_{N}=0} = (x_{1}, ...,\, x_{N-1},0)$ into the source term on the RHS of the inhomogeneous generalized diffusion equation (\ref{fn-1EQ}). Then the inhomogeneous equation (\ref{fn-1EQ}), supplemented by Eqs.\ (\ref{n-1NC}), (\ref{n-1RC}), $f^{(N-1)}(\vec{x}_{N-1};0)=0$ constitutes again a closed
initial-boundary value problem. Having obtained its solution, the partial PDF $f^{(N-1)}(\vec{x}_{N-1};t)$, we again evaluate it at $\left.\vec{x}_{N-1}\right|_{x_{N-1}=0}$
and use as the source term in the equation for partial PDF $f^{(N-2)}(\vec{x}_{N-2};t)$. The hierarchy is terminated by the Eq.\ (\ref{f0EQ}).
\subsection{\label{subsec:initialconditions} Initial distribution of particles}

Among all possible forms of initial distributions there exists one which leads to a considerable simplification of the subsequent analytical calculations. Even more importantly, the distribution is physically quite natural for real single-file systems. It describes the state which emerges, e.g., after previous constitution of a steady state\cite{RC2011},
or after previous autonomous relaxation towards a thermodynamic equilibrium. Namely, in this paper we consider the initial joint PDF for the $N$-particle system in the form
\begin{equation}
\label{iniGeneral}
	f^{(N)}(\vec{x}_{N};0) = \left\{
	\begin{array}{ll}
	 		\displaystyle N!\prod_{n=1}^{N} f(x_{n})
			& \,\, \textrm{for} \,\, \vec{x}_{N} \in \mathcal{R}_{N} \,\,,\\[15pt]
			0
			& \,\, \textrm{for} \,\, \vec{x}_{N} \notin \mathcal{R}_{N} \,\,,
	\end{array}
	\right.
\end{equation}
where $\mathcal{R}_{N}$ denotes a \emph{phase space} defined as a space of all possible configurations of $N$ hard-core interacting particles on the semi-infinite interval $[0,+\infty)$, i.e.,
\begin{equation}
\label{PhaseSpace}
\mathcal{R}_{N}=\left\{\vec{x}_{N}: 0\leq x_{N}<x_{N\!-1}< ... < x_{1}< + \infty\right\}\,\,.
\end{equation}
The symbol $\vec{x}_{N} \in \mathcal{R}_{N}$ means that the components  of the vector $\vec{x}_{N}=(x_{1},...\,,x_{N})$ respect all the inequalities in (\ref{PhaseSpace}).
If $\vec{x}_{N} \notin \mathcal{R}_{N}$, then at least one of them is violated.

In general, the function $f(x)$ in Eq.\ (\ref{iniGeneral}) can be any PDF defined and normalized on the half-line $x\in [0,+\infty)$. The normalization of the function $f(x)$ implies the proper normalization of the initial condition (\ref{iniGeneral}). This can be checked by applying Eq.\ (\ref{integrace}) from Appendix\ \ref{phasespaceint} where we have
summarized the basic rules concerning the integration over the phase space $\mathcal{R}_{N}$.

In the following, we derive also the {\em marginal\/} PDF for the position of the $n$-th particle. At the time $t=0$, it is obtained by integrating
the function (\ref{iniGeneral}) over the coordinates of all other $N-1$ particles. If $\psi_{n}^{(N)}(x_{n};0)$ denotes the marginal PDF in question, we have
\begin{eqnarray}
\label{inipsi}
\psi_{n}^{(N)}(x_{n};0) &=&   \int\limits_{\mathcal{R}_{N}\backslash x_{n}}\!\!
[{\rm d} \vec{x}_{N} \backslash x_{n}]
f^{(N)}(\vec{x}_{N};0)=\\
\notag
 &=&  \frac{N!}{N_{L}!N_{R}!} f(x_{n})\left[ \int_{0}^{x_{n}}\!\!\!\!\!\!\! {\rm d}x \, f(x) \right]^{N_{L}}\!
\left[ \int_{x_{n}}^{+\infty}\!\!\!\!\!\!\!\!\! {\rm d}x\, f(x) \right]^{N_{R}}\!\! ,
\end{eqnarray}
where $N_{L} = N-n$ ($N_{R} = n-1$) stands for the number of particles located left (right) of the $n$-th particle. The symbol $[{\rm d}\vec{x}_{N} \backslash x_{n}]$ is explained in Eq.\ (\ref{symbol}) from Appendix\ \ref{phasespaceint} and the symbol ${\mathcal{R}_{N} \backslash x_{n}}$, designating the integration domain, is explained in the paragraph just above Eq.\ (\ref{symbol}).

The last expression has been written in the form which suggests two important interpretations. First, suppose that we have generated $N$ points on the half-line, each of them being
independently drawn from the distribution with PDF $f(x)$. Then $\psi_{n}^{(N)}(x;0)$ is the PDF for the $n$-th rightmost of them (i.e., for the $n$-th order statistics\cite{karlin}).
Second, consider the system of $N$ identical, {\em indistinguishable, and noninteracting\/} particles. Let the position of each of them be described by PDF $f(x)$. Then
$\psi_{n}^{(N)}(x;0) {\rm d}x$ equals the probability that there is a particle in $(x,x+{\rm d}x)$ and, simultaneously, there are precisely $N_{L}$ ($N_{R}$) particles at arbitrary coordinates to the left (to the right) of it.

In the following, provided we need a further specification of PDF $f(x)$, we always take
\begin{equation}
\label{definicef}
f(x) = \frac{1}{L}\exp{\!\left(- \frac{x}{L} \right)}\,\,, \qquad L > 0\,\,.
\end{equation}
This expression together with Eq.\ (\ref{iniGeneral}) yields a complete specification of the initial state. The state can be prepared in the following way. Up to the time $t=0$,
a) the pore is closed (reflecting boundary at the left end of the pore at $x=0$), b) each particle is pushed towards the reflecting boundary by the external space-homogeneous force
(e.g.\ the particles are charged and we assume an external electric field), c) the $N$-particle system has reached the state of thermal equilibrium.
At the time $t=0$ instantaneously (adiabatically) a) the pore becomes opened, i.e., we replace the reflecting boundary by the partially absorbing one, b) the external field is switched off.
\subsection{\label{subsec:mapping} Mapping on single-particle diffusion in  N dimensions}

\begin{figure}[h!]
\includegraphics[scale=0.8]{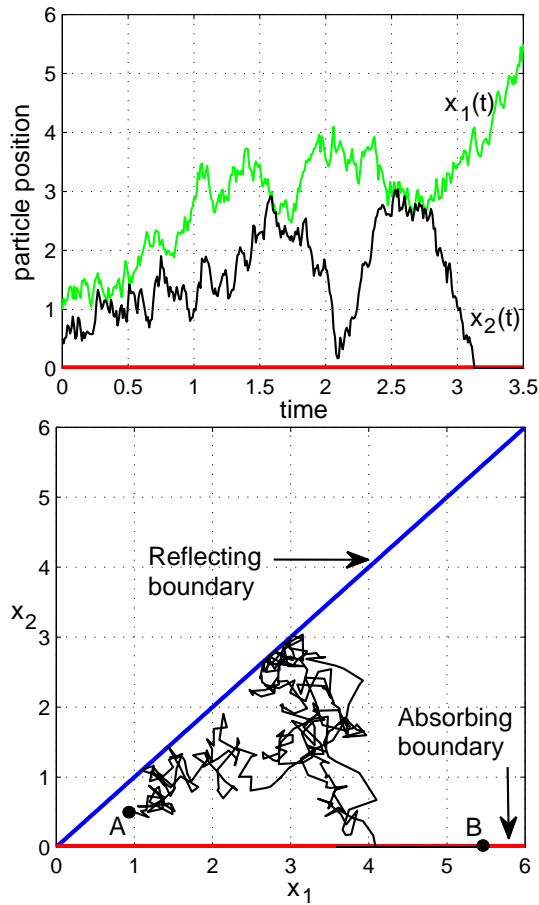}
\caption{Mapping of the two-particle SFD onto the diffusion of one compound particle in the 2D wedge domain. The upper panel shows the space-time representation of the diffusion within the pore. The red line $x=0$ is the absorbing boundary. Whenever the particles collide, their trajectories are mutually reflected. One such event occurs roughly at the time $t\approx 1.6$. The second particle leaves the pore approximately at the time $t\approx 3.2$. At the moment, the coordinate of the first particle is
roughly $x_{1}\approx 4.1$. The lower panel presents the diffusion of the compound particle within the wedge domain bounded by the reflecting boundary (the blue line) and by the absorbing boundary (the red line). The point ${\rm A}\!=\![1,0.5]$ shows the initial position of the compound particle. Whenever its trajectory hits the blue line it is reflected. The reflection corresponds to the collisions of the two particles. At the point $[x_{1},x_{2}]\approx[4.1,0]$, the trajectory of the compound particle sticks to the absorbing boundary. This represents the exit of the second particle from the pore at the time $t\approx 3.2$. Afterwards, the compound particle continues to diffuse along the line $x_{2}=0$. Its final position at the time $t=3.5$ is shown by the point ${\rm B}\!\approx\![5.5,0]$.}
\label{fig:wedgedomain}
\end{figure}

The dynamics described in the previous subsection possess a noteworthy geometric interpretation. It resides in mapping of the $N$ interacting diffusing particles onto a single ``compound'' particle in $N$ dimensions.
At $t=0$ there are $N$ particles inside the pore. Until the exit of the $N$-th particle from the pore, the situation corresponds to the diffusion of the compound particle in the $N$ dimensional wedge domain
$$\left\{\vec{x}_{N}:  0 \leq x_{N} < x_{N\!-1} < ... <x_{2} < x_{1} < + \infty  \right\}$$
bounded by $N-1$ reflecting hyperplanes
$$\left\{\vec{x}_{N}:0 \leq x_{N} < ...<x_{n}=x_{n-1}<...<x_{1}< + \infty  \right\}\,\,,$$
$n=2,...,N$, and by the absorbing hyperplane
$$\left\{\vec{x}_{N}: x_{N}=0, 0 \leq x_{N\!-1} < ... <x_{2} < x_{1} < + \infty  \right\}\,\,.$$
The diffusion within this wedge is described by the partial PDF $f^{(N)}(\vec{x}_{N};t)$ and it is governed by the $N$-dimensional diffusion equation (\ref{fnEQ}).
In this picture, the non-crossing boundary conditions (\ref{nNC}) describe the reflection of the compound particle on $N-1$ reflecting hyperplanes. They guarantee
the zero probability currents in the directions perpendicular to these hyperplanes. The exit of the left-most particle from the pore corresponds to ``sticking'' of the compound particle to the absorbing hyperplane and it is incorporated by the requirement (\ref{nRC}).

The exit of the $N$-th particle translates to the further diffusion of the compound particle on the above absorbing hyperplane. Equivalently, it diffuses within the $N\!-\!1$-dimensional wedge domain
$$\left\{\vec{x}_{N-1}:  0 \leq x_{N\!-1} < ... <x_{2} < x_{1} < + \infty  \right\}$$
bounded by $N-2$ reflecting hyperplanes
$$\left\{\vec{x}_{N-1}:0 \leq x_{N-1} < ...<x_{n}=x_{n-1}<...<x_{1}< + \infty  \right\}\,\,,$$
$n=2,...,N-1$, and by the absorbing hyperplane
$$\left\{\vec{x}_{N-1}: x_{N-1}=0, 0 \leq x_{N\!-2} < ... <x_{2} < x_{1} < + \infty  \right\}\,\,.$$
The diffusion within this domain is described by the partial PDF $f^{(N-1)}(\vec{x}_{N-1};t)$, etc.

Eventually, after the exit of the $N-2$ particles, there are two last particles left inside the pore. This case is illustrated in Fig.\ \ref{fig:wedgedomain} where the compound particle diffuses in 2D wedge domain $\left\{\vec{x}_{2}:  0 \leq x_{2} < x_{1} < + \infty  \right\}$ bounded by just one reflecting boundary $\left\{\vec{x}_{2}:0\leq x_{2}=x_{1}< + \infty  \right\}$ and by the absorbing half-line $\left\{\vec{x}_{2}: x_{2}=0, 0\leq x_{1} < + \infty  \right\}$. Its motion is described by the partial PDF $f^{(2)}(\vec{x}_{2};t)$.
After this event, the diffusion of the compound particle is described by partial PDF $f^{(1)}(x_{1};t)$. The exit of the left particle corresponds to the sticking of the compound particle to the half-line $\{\vec{x}_{2}: x_{2}=0, 0\leq x_{1}<+\infty \}$. The exit of the last particle simply means the sticking of the compound particle to the origin. The probability of finding the compound particle at the origin at $t$ is given by the function $f^{(0)}(t)$.
For a further discussion of the 2D case cf.\ Fig.\ \ref{fig:wedgedomain}.

\section{\label{sec:solution}Exact solution of Evolution Equations}
\subsection{\label{subsec:partialPDFs}Partial PDFs}
In our recent work we have shown that the Green function (GF) for the diffusion equation of $N$ hard-core interacting particles diffusing in an {\em arbitrary\/} external potential can be constructed from the GF $p(x;t\,|\,y;t')$ of the corresponding single-particle problem (see Eq.\ (6) in Ref.\ \onlinecite{RC2011}).
In the present context, the same construction yields GF that solves Eq.\ (\ref{fnEQ}) with the non-crossing boundary condition (\ref{nNC}), with the radiation boundary condition (\ref{nRC}), and with the initial condition $p^{(N)}(\vec{x}_{N};t'\,|\,\vec{y}_{N};t') = \delta(\vec{x}_{N}-\vec{y}_{N})$. The explicit form of this GF is given by
the permanent\cite{permanents}
\begin{eqnarray}
\notag
 p^{(N)}(\vec{x}_{N}; t \, | \, \vec{y}_{N}; t') =
\sum_{m=1}^{N!} p(x_{1}; t \, | \, y_{\pi_{m}(1)}; t')
 p(x_{2}; t \, | \, y_{\pi_{m}(2)}; t') ...\,
p(x_{N}; t \, | \, y_{\pi_{m}(N)}; t')\,\,,
\label{perminitial}
\end{eqnarray}
if $\vec{x}_{N}, \vec{y}_{N} \in \mathcal{R}_{N}$, and it vanishes otherwise. The summation runs over all $N!$ permutations of initial positions. Having the GF, the initial condition (\ref{iniGeneral}) evolves as
\begin{equation}
f^{(N)}\!(\vec{x}_{N};t) = \!\!
\int\limits_{\mathcal{R}_{N}} \!\!{\rm d}\vec{y}_{N} \,
p^{(N)}(\vec{x}_{N};t\,|\,\vec{y}_{N};0) f^{(N)}(\vec{y}_{N};0)\,\,.
\label{PNXT}
\end{equation}
In the present paper we always assume the initial condition (\ref{iniGeneral}) and it is obvious that this special form simplifies the analytical expressions.
As a matter of fact, using (\ref{iniGeneral}), the above function reduces to the simple product (cf.\ Appendix \ref{app:permanent})
\begin{equation}
\label{PNXTproduct}
f^{(N)}(\vec{x}_{N};t) = N!\, p(x_{N}; t) p(x_{N-1}; t) ... p(x_{1}; t)\,\,,
\end{equation}
where $p(x;t)$ is the known solution of the corresponding single-particle problem (cf. Eq.\ (\ref{pDEF})). Notice that the product yields the solution just within the domain
$\mathcal{R}_{N}$. It turns out that the products of this type will play an important role in the following. Hence we define
\begin{equation}
\label{PNprodDEF}
	p^{(N)}(\vec{x}_{N};t)=\left\{
	\begin{array}{ll}
	 		\displaystyle N!\prod_{n=1}^{N} \, p(x_{n}; t)
			& \,\, \textrm{for} \,\, \vec{x}_{N} \in \mathcal{R}_{N} \,\,,\\[15pt]
			0
			& \,\, \textrm{for} \,\, \vec{x}_{N} \notin \mathcal{R}_{N} \,\,,
	\end{array}
	\right.
\end{equation}
We are now prepared to formulate the exact solution of our diffusion problem.

{\em
Suppose that the initial PDF $f^{(N)}(\vec{x}_{N};0)$ possess the form {\rm (\ref{iniGeneral})}. Then the hierarchy of coupled evolution equations as
formulated in {\rm Subsec.\ \ref{subsec:evolutioneqs}} possesses the solution
\begin{equation}
\label{theorem1}
f^{(N\! -k)}(\vec{x}_{N\! -k};t) \! =\!\! \binom{N}{k} p^{(N\! -k)}(\vec{x}_{N\! -k};t)\!\!
\left[ \kappa\!\!\int_{0}^{t}\!\!\! {\rm d}t' p(0;t')\right]^{k}\! ,
\end{equation}
for $k = 0, 1, ...\,, N-1$, and
\begin{equation}
\label{theorem2}
f^{(0)}(t) = \, \left[ \kappa\!\!\int_{0}^{t}\!\! {\rm d}t' p(0;t') \right]^{N}.
\end{equation}
\/}
Turning to the proof, the solution of the first element of the hierarchy (the case $k=0$) has already been derived above. Actually, provided there are still $N$ particles in the pore, the solution (\ref{theorem1}) coincides with Eq.\ (\ref{PNXTproduct}).

For an arbitrary fixed $k=1,\ldots,N-1$, we again first consider the $(N\!-\!k)$-particle GF that solves Eq.\ (\ref{fn-kEQ}) with the non-crossing BC (\ref{n-kNC}), with the radiation BC (\ref{n-kRC}), and with the initial condition $p^{(N-k)}(\vec{x}_{N-k};t'\,|\,\vec{y}_{N-k};t') = \delta(\vec{x}_{N-k}-\vec{y}_{N-k})$. It reads
\begin{equation}
 p^{(N-k)}(\vec{x}_{N-k}; t \, | \, \vec{y}_{N-k}; t') =
\sum_{m=1}^{(N-k)!} p(x_{1}; t \, | \, y_{\pi_{m}(1)}; t')
 p(x_{2}; t \, | \, y_{\pi_{m}(2)}; t') ...\,
p(x_{N-k}; t \, | \, y_{\pi_{m}(N-k)}; t')\,,
\label{perminitialn-k}
\end{equation}
if $\vec{x}_{N-k}, \vec{y}_{N-k} \in \mathcal{R}_{N-k}$, and it vanishes otherwise.

In order to derive the partial PDF $f^{(N-k)}\!(\vec{x}_{N-k};t)$ we first use the above GF (\ref{perminitialn-k}) to evolve the initial condition $f^{(N-k)}\!(\vec{x}_{N-k};0)$. This yields the first term on the right-hand side of the following equation. Secondly, we take into account the source term in Eq.\ (\ref{fn-kEQ}) which gives the second term on the right-hand side of the following equation. Altogether, we obtain
\begin{eqnarray}
f^{(N-k)}\!(\vec{x}_{N-k};t) &=& \!\!
\int\limits_{\mathcal{R}_{N-k}} \!\!{\rm d}\vec{y}_{N-k} \,
p^{(N-k)}(\vec{x}_{N-k};t\,|\,\vec{y}_{N-k};0) f^{(N-k)}(\vec{y}_{N-k};0)+ \\
&&+\int\limits_{\mathcal{R}_{N-k}} \!\!{\rm d}\vec{y}_{N-k} \,
\int_{0}^{t} \!\!{\rm d}t' \,
p^{(N-k)}(\vec{x}_{N-k};t\,|\,\vec{y}_{N-k};t')\, \kappa \left. f^{(N-k+1)}(\vec{y}_{N-k+1};t')\right|_{y_{N-k+1}=0}.
\notag
\label{soln-k}
\end{eqnarray}
The first term on the right-hand side vanishes because $f^{(N-k)}\!(\vec{x}_{N-k};0)=0$ (initially, there are $N$ particles in the pore).
As for the second term, we start with $k\!=\!1$ and we invoke the same reasoning which enables to rewrite the Eq.\ (\ref{PNXT}) in the simple product form (\ref{PNXTproduct}). More explicitly, inserting the product
form (\ref{PNXTproduct}) with $x_{N}=0$ into Eq.\ (\ref{soln-k}) and repeating the steps as described in Appendix \ref{app:permanent}, we carry out the required integration over the space variables. After these steps we get
\begin{equation}
f^{(N-1)}(\vec{x}_{N-1};t) =
N p^{(N-1)}(\vec{x}_{N-1};t)\kappa\!\!\int_{0}^{t}\!\!\! {\rm d}t' p(0;t')\,\,.
\label{N-1PDFproofF}
\end{equation}
This is just Eq.\ (\ref{theorem1}) for $k\!=\!1$. The procedure can be simply repeated for $k\!=\!2$. After introducing Eq.\ (\ref{N-1PDFproofF}) into Eq.\ (\ref{soln-k}) we have
\begin{equation}
f^{(N-2)}(\vec{x}_{N-2};t)  =
N(N-1) p^{(N-2)}(\vec{x}_{N-2};t) 
\kappa\!\!\int_{0}^{t}\!\!\!{\rm d}t' p(0;t')\,\kappa\!\!\int_{0}^{t'}\!\!\!\!{\rm d}t'' p(0;t'')\,\,,
\label{N-2PDFproofF}
\end{equation}
which equals (cf. Eqs. (\ref{integraceG}, \ref{integrace}))
\begin{equation}
f^{(N-2)}(\vec{x}_{N-2};t) =
\frac{N(N-1)}{2!} p^{(N-2)}(\vec{x}_{N-2};t) 
 \left[\kappa\!\!\int_{0}^{t}\!\!\!{\rm d}t'p(0;t')\right]^{2}\,\,.
\label{N-2PDFproofIIF}
\end{equation}
This proves the statement (\ref{theorem1}) for $k\!=2\!$. The proof is completed by repeating the procedure for $k=3,\ldots,N$.
\subsection{\label{subsec:taggedPDFs}Tagged particle's PDF}

In the preceding section, we have derived the probabilities $f^{(N-k)}(\vec{x}_{N-k};t){\rm d}\vec{x}_{N-k}$ of the compound event that, at the time $t$, the pore contains $N\!-\!k$ particles \emph{and} that they are found in $(\vec{x}_{N-k}, \vec{x}_{N-k}+{\rm d} \vec{x}_{N-k})$. We now deduce the properties of the single, say $n$-th particle (we shall call it the $n$-th tagged particle). We start from the partial PDFs $f^{(N-k)}(\vec{x}_{N-k};t)$, we integrate out the coordinates of all other particles \emph{and} we sum over all $k$.

The probability that, at the time $t$, the pore contains $N-k$ particles and the $n$-th particle is found in the infinitesimal vicinity of the coordinate $x_{n}$ is
$\psi_{n}^{(N-k)}(x_{n};t)\,{\rm d}x_{n}$, where (for the definition of the symbols used cf.\ the text after Eq.\ (\ref{inipsi}) and the last paragraph in the Appendix\ \ref{phasespaceint})
\begin{equation}
\label{tagged}
\psi_{n}^{(N-k)}(x_{n};t)=\!\!\!\!\!\!\!\!\!
\int\limits_{\mathcal{R}_{N-k}\backslash x_{n}}
\!\!\!\!\!\!
[{\rm d} \vec{x}_{N-k} \backslash x_{n}]
f^{(N-k)}(\vec{x}_{N-k};t)\,\,.
\end{equation}
To find the $n$-th particle in the pore, the pore must contain at least $n$ particles. Hence the marginal PDF of the $n$-th particle's position regardless the positions of all
other particles \emph{and} regardless the number of the particles in the pore is given by
\begin{equation}
\psi_{n}(x;t) = \sum_{k=0}^{N-n} \psi_{n}^{(N-k)}(x;t)\,\,.
\end{equation}
We now insert the partial PDFs as given by (\ref{theorem1}) into Eq.\ (\ref{tagged}). Skipping out a considerable amount of purely algebraic steps, the PDF for the $n$-th tagged particle emerges in the form
\begin{equation}
\psi_{n}(x;t) = \frac{N!}{(n-1)!(N-n)!} p(x;t)
\left[ \int_{x}^{+\infty}\!\!\!\! {\rm d}x'p(x';t) \right]^{n-1} 
 \left[1-S(t)+F(x,t) \right]^{N-n}.
\label{marginalf}
\end{equation}
Notice that, as promised above, the result is formulated solely by means of the single-particle functions $p(x;t)$, $S(t)$, and $F(x,t)$, as defined in Eqs.\ (\ref{pDEF}), (\ref{FDEF}), and (\ref{survprob}), respectively. In closing the calculation we would like to emphasize the relation
\begin{equation}
\sum_{n=1}^{N} \psi_{n}(x;t) =Np(x;t)\,\,
\label{sumdens}
\end{equation}
which can proved by the direct summation of the expressions (\ref{marginalf}). The result is also plausible if we apply the reasoning elaborated in the following paragraph.

In the text following Eq.\ (\ref{inipsi}) we have emphasized the analogies of our SFD model with a) the system of $N$ noninteracting indistinguishable particles,
and with b) the distribution of the $n$-th order statistics based on the sample of $N$ independent, identically distributed random variables. However, in Subsec.\ \ref{subsec:initialconditions}, the discussion refers to the initial distribution of the particles. Eq.\ (\ref{marginalf}) testifies that the analogies are valid
for an arbitrary time if we only supplement them with the necessary modifications stemming from the escape process. Indeed, $\psi_{n}(x;t){\rm d}x$ gives the simultaneous
probability that one of the $N$ non-interacting indistinguishable particles is found in $(x,x+{\rm d}x)$, $n-1$ particles are located in $(x,+\infty)$, and each of the
remaining $N-n$ particles is either already removed from the pore (with the probability $1-S(t)$) or it is still diffusing in the domain in $(0,x)$ (with the probability
$F(x,t)$). The indistinguishability is behind the factorials. They account for all possible sequences of particles to the right and to the left of the tagged
particle. On the other hand, consider $N$ independent, identically distributed random variables each one possessing the probability density $p(x;t)+(1-S(t))\,\delta(x)$. Then
Eq.\ (\ref{marginalf}) is just the probability density for the $n$-th biggest among them. In this second analogy, the time enters as a parameter.
\section{\label{sec:escapeprocess}Escape process}
\subsection{\label{subsec:survival}Survival probabilities}
If we use the initial condition (\ref{definicef}) and if we carry out the steps as given in Subsec.\ \ref{subsec:singleparticle}, the single-particle survival probability (\ref{survprob}) reads
\begin{eqnarray}
S(t) =  \frac{1}{\kappa L - D} && \left[  \kappa L
\exp{\! \left(\frac{Dt}{L^{2}} \right)}\, {\rm erfc}\!\left(\frac{\sqrt{Dt}}{L}\right) -   
D \exp{\! \left(\kappa^{2}\frac{t}{D} \right)}\, {\rm erfc}\!\left( \kappa \sqrt{\frac{t}{D}} \right) \right] \,\,.
\label{S}
\end{eqnarray}
At $t=0$, the survival probability equals to one. It approaches zero as $t \rightarrow \infty$, i.e., the particle will certainly exit the pore.
The survival probability (\ref{S}) decays with time as the power law\cite{Note1}
\begin{equation}
\label{Sdecay}
S(t) \sim A\, t^{-\frac{1}{2}}\,\,,\quad \mathrm{as} \quad t \rightarrow \infty\,\,.
\end{equation}
This asymptotic time dependence is determined solely by the geometry of the problem (1D semi-infinite interval with the absorbing boundary). The constant factor $A$ includes the parameters $L, D, \kappa$. It reads
\begin{equation}
A = \frac{1}{\sqrt{\pi D}}\left( L+\frac{D}{\kappa} \right) \,\,.
\end{equation}
We shall demonstrate that the parameters $L$, $D$, $\kappa$ enter the time-asymptotic behavior of the interacting particles only through this combination.
\begin{figure}[t]
\includegraphics[scale=1]{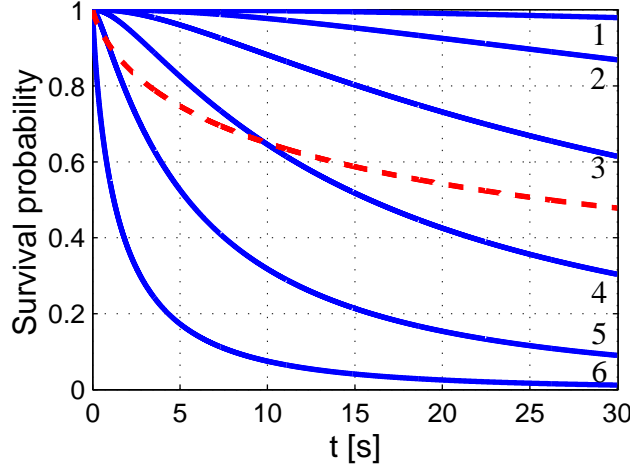}
\caption{The survival probabilities $S_{n}(t)$ for the tagged particles (Eq.\ (\ref{survn}), solid blue lines), and the single-particle survival probability $S(t)$ (Eq.\ (\ref{S}), the dashed red line) as the functions of time. The parameters used: $N=6$, $L=5\,{\rm m}$, $D=1\,{\rm m^{2}s^{-1}}$, $\kappa=1\,{\rm ms^{-1}}$. The number $n = 1,...\,,6$ on the right near the blue lines denotes the order of the corresponding particle, cf.\ Fig.\ 1.}
\label{fig:survival}
\end{figure}

Having prepared the single-particle survival probability $S(t)$ we now turn to the system of interacting particles. The survival probability $S_{n}(t)$ of the $n$-th tagged particle
is defined as
\begin{equation}
\label{Sndefin}
S_{n}(t) = \int_{0}^{+\infty}\!\!\!\!\!\!{\rm d} x\, \psi_{n}(x;t)\,\,,\,\,n=1,2,\ldots,N\,\,,
\end{equation}
where $\psi_{n}(x;t)$ is the $n$-th particle's PDF as given by Eq.\ (\ref{marginalf}). The main steps of the emerging calculation are summarized in Appendix \ref{app:SPC}.
The resulting expression includes only the single-particle survival probability:
\begin{equation}
 S_{n}(t) = n \binom{N}{n}\!   \left(S(t)\right)^{n}
  \sum_{k=0}^{N-n}\frac{(-1)^{k}}{n+k} \binom{N-n}{k} \left(S(t)\right)^{k} \,\,.
\label{survn}
\end{equation}
This is the main formula of the present section.

To get a better insight into the escape process in question, let us examine the difference between the survival probabilities of the adjacent particles. We have
\begin{equation}
 S_{n}(t)-S_{n+1}(t) =  \binom{N}{n}\!   \left(S(t)\right)^{n}  \left(1-S(t)\right)^{N-n} \,\,.
\label{rozdilS}
\end{equation}
The difference is simply the probability that the particle No.\ $n+1$ has already exited the pore and, simultaneously, the particle No.\ $n$ is still diffusing inside the pore.
At the same time, the right-hand side by itself suggests an interpretation in terms of a system of \emph{noninteracting and indistinguishable particles} (the  survival properties of $N$ noninteracting particles are discussed e.g. in Refs.\ \onlinecite{Lindenberg, Yuste, Krapivsky}). Eq.\ (\ref{rozdilS}) tells us that the following two events have the same probability: (i) exactly $N-n$ of the initial number $N$ of interacting particles have already exited the pore by the time $t$, and (ii) exactly $N-n$ of the initial number $N$ of noninteracting and indistinguishable particles have already exited the pore by the time $t$. Differently speaking, the mass transport out of the pore is not affected by the hard-core interaction. In particular, in the probabilistic sense, it will take the same time until all $N$ particles escape from the pore regardless they interact or not.

Fig.\ \ref{fig:survival} illustrates the tagged particle survival probability $S_{n}(t)$ for different $n = 1,...\,,6$, together with the single-particle survival probability $S(t)$.
Several observations are worth emphasizing. First of all, it follows from Eq.\ (\ref{rozdilS}) that the function $S_{n}(t)$ are ordered as
\begin{equation}
S_{1}(t) \geq S_{2}(t) \geq ... \geq S_{N}(t)\,\,,
\label{nerovnosti}
\end{equation}
where the equalities hold at $t=0$ (all $N$ particles are initially inside the pore, thus all $S_{n}(0)=1$) and in the limit $t \to \infty$. Each hard-core particle acts on the adjacent ones as a moving (fluctuating) reflecting boundary. For example, the exit of the rightmost particle is hindered by the remaining $N\!-\!1$ particles which greatly increases its survival probability $S_{1}(t)$. Effectively, the collisions between the particles induce {\em entropic repulsive forces\/}. The leftmost particle is pushed by the remaining ones towards the pore's boundary which significantly reduces its lifetime inside the pore; its survival probability $S_{N}(t)$ rapidly decreases to zero. As a direct consequence of Eq.\ (\ref{sumdens}) we obtain
\begin{equation}
\frac{1}{N}\sum_{n=1}^{N} S_{n}(t)=S(t) \,\,
\end{equation}
which relates the arithmetic mean of tagged particles' survival probabilities to the survival probability of the single-diffusing particle. Hence $S_{1}(t)$ always exceeds
$S(t)$ but their difference $S_{1}(t)\!-\!S(t)$ vanishes as $t\to \infty$.

The above consideration manifest themselves most pronouncedly if we consider the time-asymptotic behavior of the survival probabilities. Each $S_{n}(t)$ tends to zero, i.e., each particle will certainly exit the pore. The long-time asymptotics of the individual $S_{n}(t)$ can be obtained directly from Eq.\ (\ref{survn}). We have
\begin{equation}
\label{Snasymptotics}
 S_{n}(t)\,  \sim \binom{N}{n} A^{n}\, t^{-\frac{n}{2}}  \,\,, \quad \mathrm{as} \quad t \rightarrow \infty\,\,,
\end{equation}
where $A$ has the same meaning as in Eq.\ (\ref{Sdecay}). In words, each survival probabilities exhibits again a power-law decay. However, the decay exponent depends on the particle's order $n$ and it {\em does not\/} depend on the total number of particles. The parameter $N$ enters just through the constant prefactor of the power law. This can be understood on physical grounds. Notice that the survival probability of the rightmost particle $S_{1}(t)$ decays as $t^{-1/2}$, i.e., with the same dynamical exponent as $S(t)$ (cf. Eq.\ (\ref{Sdecay})). Indeed, in the long-time limit it is highly probable that all other particles have already escaped and thus the initially rightmost particle behaves as a free one.
This is also why the total number of particles $N$ does not enter the exponents of $S_{n}(t)$, for an arbitrary $n$.
After a ``long enough'' time, the $N\!-\!n$ particles which were initially to the left of the $n$-th one have already escaped. Thus, irrespective of $N$, the $n$-th particle behaves as if it were the leftmost one in the system of $n$ particles. The total number of particles $N$ specifies only the expression ``long enough'' in the above reasoning.

Finally, let us return to the mapping which has been developed in Subsec.\ \ref{subsec:mapping} where we have discussed the analogy between our SFD problem and the diffusion of a single compound particle in the $N$-dimensional space. Initially, the compound particle departs from a general interior point in the $N$-dimensional wedge domain. The exit of the
leftmost particle has been translated to the adsorption of the compound particle onto the $(N\!-\!1)$-dimensional absorbtion hyperplane. Then the compound particle continues to diffuse within this hyperplane until its motion is further restricted onto the $(N\!-\!2)$-dimensional absorbtion hyperplane, and so on. The function $S_{n}(t)$ equals the probability that, at the time $t$, the motion of the compound particle has not been restricted onto the $(n\!-\!1)$-dimensional hyperplane, yet. That is, it still moves in some $k$-dimensional wedge, where $k\geq n$. Similarly, the difference $S_{n}(t)-S_{n+1}(t)$ as given in Eq.\ (\ref{rozdilS}) yields the probability that, at the time $t$, the compound particle diffuses in the interior of the $n$-dimensional wedge domain. On the whole, as a byproduct, we have solved a nontrivial problem concerning the geometrically restricted $N$-dimensional diffusion.
\subsection{\label{subsec:escapetime}Exit times}
The survival probability $S_{n}(t)$ (Eq.\ (\ref{survn})) provides complete information about the escape process of the $n$-th particle. The PDF $\phi_{n}(t)$ of the $n$-th particle's \emph{exit time} is given by\cite{Redner}
\begin{equation}
\phi_{n}(t) = - \frac{ {\rm d} S_{n}(t)}{ {\rm d} t}\,\,.
\end{equation}
Hence all the moments of the exit time can be obtained from $S_{n}(t)$. It follows from the long-time properties of $S_{n}(t)$ (cf.\ Eq.\ (\ref{Snasymptotics})) that
\begin{equation}
\label{exitPDF}
 \phi_{n}(t)\,  \sim \frac{n}{2} \binom{N}{n} A^{n}\,  t^{-\frac{n}{2}-1}  \,\,,
 \quad \mathrm{as} \quad t \rightarrow \infty\,\,.
\end{equation}
Consequently, the mean exit time of the $n$-th particle,
\begin{equation}
\label{escapetime}
\left< t_{n} \right>  = \int_{0}^{+\infty}\!\!\!\!\!\! {\rm d} t\,\, t\, \phi_{n}(t)\,\,,
\end{equation}
is infinite for $n \leq 2$. The integral (\ref{escapetime}) becomes finite for $n>2$ and its convergence does not depend on the total number of particles initially located in the pore. $N$ only modifies the magnitude of the mean exit time (if the latter exists). Similarly, the second moment of the exit time $\left<t_{n}^{2}\right>$ only exists for
$n>3$.

The above analysis brings us to an interesting comparison between the escape process confined by the static and by the fluctuating reflecting boundaries, respectively. Considering the semi-infinite interval with the absorbing boundary at the origin, the survival probability $S(t)$ of the single-diffusing particle decays to zero as the power law (\ref{Sdecay}) and the mean exit time is infinite. If we add the \emph{static} reflecting boundary on the right of the single-diffusing particle, the diffusion will be confined to an interval of a finite length. Then the survival probability decays exponentially and, consequently, the mean exit time becomes finite\cite{Redner}. Consider now our SFD model with $N=2$. Then, for the left particle, the right one plays the role of a moving reflection boundary. That is, the left particle diffuses within a finite interval of a fluctuating length. Contrary to the case of the static boundary, the survival probability of the left particle decays as $1/t$, and its mean exit time is infinite. It is interesting that the case $N=3$ brings us to still another behavior of the leftmost particle. For this particle, the central particle again represents a moving boundary. This boundary itself diffuses and, moreover, it feels another moving boundary to the right of it. As a result of such right-hand confinement the escape process of the leftmost particle is accelerated, its mean exit time is finite, but the variance of the exit time still diverges. For a general $N$, the exit time of any $n$-th tagged particle with $n\geq 4$ possesses both a finite mean value and a finite variance.

In leaving the subject, notice that the limit $\kappa \to \infty$ alters only the prefactor $A$ but not the long-time decay exponent of $S_{n}(t)$. This limit corresponds to perfect absorption (particle is absorbed immediately when it hits the absorbing boundary). Consequently, $\displaystyle \lim_{\kappa \to \infty} \phi_{n}(t)$ gives the distribution of the $n$-th particle's \emph{first-passage time} and $\displaystyle \lim_{\kappa \to \infty} \left< t_{n} \right>$  equals the mean first-passage time.
\section{\label{sec:outlook}Conclusion}
This paper provides a simple and general formalism for 1D SFD problems with absorbing boundaries. Our main message concerns the relation of the physically motivated SFD problem and a broadly developed probabilistic field of order statistics and it can be summarized as follows. If the initial condition in the physical SFD problems is assumed in the form of a joint PDF for the order statistics based on the system of identical and independent parent random variables, then this form is preserved during the subsequent time evolution. Implicitly, the observation has been already contained in the previous theoretical SFD studies in problems without boundaries. It can be easily read from the permanent-like structure of the Green function.

Our work elaborates the connection between the SFD problem and the order-statistics point of view into new directions. First of all, we have proven that the relation continues to be valid in the problems with boundaries. In these problems, the arbitrary-time joint PDF for the positions of the particles has again the form of the joint PDF for the order statistics, provided one simply admits that the PDF for a single parent random variable has an atom at the origin. Secondly, specifically in the problems with boundaries, the joint PDF for the order statistics at an arbitrary time includes, in a mixed form, both the information on the particles remaining still in the pore and on their exit times. We have shown that these two pieces of information can be separated thereby yielding new exact results on the tagged particle diffusion and on the escape process, respectively.

The robustness of the relation is easily understood as it lies in both cases (with and without the boundaries) in the hard-core inter-particle interaction. It is therefore not surprising that our results are independent on a detailed form of the single-particle dynamics. They hold true, for example, if the single particle is acted upon by a
general space-dependent or even a space-and-time-dependent external force. The overall structure of our solution remains unchanged if we only change the details of the escape process and switch, for example, from the radiating boundary to the totally absorbing one. Physically motivated calculations in these directions are in progress and will be reported elsewhere.

Eventually, let us comment on our choice of the initial condition as given by Eq.\ (\ref{iniGeneral}). Obviously, it is {\em not\/} the most general option. For example, what happens if the particles are initially placed at specific set of coordinates $y_{1}<y_{2}<\ldots<y_{N}$? Taking a general initial condition, the partial PDF as defined by Eq.\ (\ref{PNXT}) {\em does not\/} reduce to the simple product form (\ref{PNXTproduct}). 
Instead, it assumes a rather involved permanent-like form similar to the RHS of Eq.\ (\ref{perminitial}). With the necessary changes, one can carry out all the steps which give the partial PDFs
$f^{(N\! -k)}(\vec{x}_{N\! -k};t)$, the tagged particle PDF $\psi_{n}(x;t)$, and the survival probabilities $S_{n}(t)$. But again, all the final expressions contain permanents. In spite of these more complicated finite-time expressions, the long-time dynamical exponents coincide with those derived in the preceding sections. Again, the parameters which specify the initial condition (together with $D$ and $\kappa$) appear in the prefactors of the power laws (\ref{Snasymptotics}), (\ref{exitPDF}).
\begin{acknowledgments}
Support of this work by the Ministry of Education of the Czech Republic (project No.\ MSM 0021620835), by the Charles University in Prague (project SSV:\ 110-09/265305) and by the Charles University Grant Agency (projects No.\ 301311, and No.\ 251228) is gratefully acknowledged.
\end{acknowledgments}
\appendix
\section{Integration over the phase space $\mathcal{R}_{N}$}
\label{phasespaceint}
Let $G(x_{1},x_{2}, ...\,,x_{N})$ be a symmetric function of its arguments, i.e., let
$$G(x_{1},x_{2}, ...\,,x_{N}) = G(x_{\pi_{m}(1)},x_{\pi_{m}(2)}, ...\,,x_{\pi_{m}(N)})\,\,$$
holds for any of $N!$ permutations  $\pi_{m}$ of the subscripts $\{1,...,\,N \}$.
Then the integration of $G(\vec{x}_{N})$ with respect to the coordinates $\vec{x}_{N}$ over the phase space $ \mathcal{R}_{N}$
(defined in Sec.\ \ref{subsec:initialconditions}) extends to the integration with respect to the components of $\vec{x}_{N}$ over the half-lines as follows,
\begin{eqnarray}
\label{integraceG}
\int\limits_{\mathcal{R}_{N}}
\!\! {\rm d}\vec{x}_{N} 
G(x_{1},x_{2}, ...\,,x_{N}) &=& 
\int_{0}^{+\infty} \!\!\!\!\!\!\!\!\!{\rm d}x_{1}\!\!
\int_{0}^{x_{1}} \!\!\!\! {\rm d}x_{2}\, ...
\int_{0}^{x_{N-1}} \!\!\!\!\!\!\!\!\!\!{\rm d}x_{N}\, G(x_{1},x_{2}, ...\,,x_{N})\\
\notag 
&=& \frac{1}{N!}
\int_{0}^{+\infty} \!\!\!\!\!\!\!\!\!{\rm d}x_{1}
\int_{0}^{+\infty} \!\!\!\!\!\!\!\!\!{\rm d}x_{2} ...
\int_{0}^{+\infty} \!\!\!\!\!\!\!\!\!{\rm d}x_{N}\, G(x_{1},x_{2}, ...\,,x_{N})\,\,.
\end{eqnarray}
The first equality introduces just another notation for the integration over $\mathcal{R}_{N}$, the second equality follows from the symmetry of $G(\vec{x}_{N})$. For a further discussion see, e.g., Appendix C in Ref.\ \onlinecite{SFDLizana}.

In particular, if $G(\vec{x}_{N})$ stands for a product of identical functions, i.e., $G(\vec{x}_{N}) = g(x_{1})g(x_{2}) ...\,g(x_{N})$, then we have
\begin{equation}
\label{integrace}
\int\limits_{\mathcal{R}_{N}}\!\! {\rm d}\vec{x}_{N}g(x_{1})g(x_{2}) ...\,g(x_{N})=
\frac{1}{N!}\left[ \int_{0}^{+\infty} \!\!\!\!\!\!\!\!\!{\rm d}x\, g(x)\right]^{N}.
\end{equation}

If we are interested in properties of a tagged particle, say of the $n$-th one, we should carry out another type of the integration over the phase space $\mathcal{R}_{N}$. In this case we have to integrate over the variables $x_{i}$, $i =1, ...,\,n-1,n+1,...,\,N$. Let us denote the domain of such integration as $\mathcal{R}_{N} \backslash x_{n}$.
Further, let us define the symbol
\begin{equation}
\label{symbol}
[{\rm d}\vec{x}_{N} \backslash x_{n}]={\rm d}x_{1} ...\, {\rm d}x_{n-1} {\rm d}x_{n+1} ...\, {\rm d}x_{N}\,\,.
\end{equation}
The integration of $G(\vec{x}_{N}) = g(x_{1})g(x_{2}) ...\,g(x_{N})$ over $\mathcal{R}_{N} \backslash x_{n}$ runs as follows
\begin{eqnarray}
\notag
\int\limits_{\mathcal{R}_{N} \backslash x_{n}}
\!\!\!\!\! [{\rm d}\vec{x}_{N} \backslash x_{n}]\,\,
g(x_{1})g(x_{2}) ...\,g(x_{N}) &=& g(x_{n})
\int_{x_{n}}^{+\infty} \!\!\!\!\!\!\!\!\!{\rm d}x_{1}\, g(x_{1})\!\!
\int_{x_{n}}^{x_{1}} \!\!\!\! {\rm d}x_{2}\, g(x_{2})  \, ...
\int_{x_{n}}^{x_{n-2}} \!\!\!\!\!\!\!\!\!\!{\rm d}x_{n-1}\,g(x_{n-1})\times \\
\notag
&&  \times \int_{0}^{x_{n}} \!\!\!\!\!\!{\rm d}x_{n+1}\,g(x_{n+1}) \,
\int_{0}^{x_{n+1}} \!\!\!\!\!\!\!\!\!\!{\rm d}x_{n+2}\,g(x_{n+2}) \, ...
\int_{0}^{x_{N-1}} \!\!\!\!\!\!\!\!\!\!{\rm d}x_{N}\,g(x_{N}) \\
\label{marginalintegrace}
&=&
\frac{g(x_{n})}{N_{L}!N_{R}!}
\left[ \int_{0}^{x_{n}} \!\!\!\!\!\!\!{\rm d}x\, g(x)\right]^{N_{L}}
\left[ \int_{x_{n}}^{+\infty} \!\!\!\!\!\!\!\!\!{\rm d}x\, g(x)\right]^{N_{R}},
\label{TPintegral}
\end{eqnarray}
where $N_{L}= N - n$ ($N_{R} = n-1$) stands for the number of particles located left (right) from the $n$-th particle
(cf.\ Fig.\ \ref{fig:singlefile}). For a more comprehensive discussion of this integration technique, see Sec. III B. in Ref.\ \onlinecite{KargerHahn}.
\section{Convolution of GF}
\label{app:permanent}
First, notice that the sum on the RHS of Eq.\ (\ref{perminitial}) can be rearranged as
\begin{eqnarray}
 p^{(N)}(\vec{x}_{N}; t \, | \, \vec{y}_{N}; t') =
\sum_{m=1}^{N!} p(x_{\pi_{m}(1)}; t \, | \, y_{1}; t')
p(x_{\pi_{m}(2)}; t \, | \, y_{2}; t') ...\,
p(x_{\pi_{m}(N)}; t \, | \, y_{N}; t')\,\,.
\label{permfinal}
\end{eqnarray}
Further, notice that $p^{(N)}(\vec{x}_{N};t\,|\,\vec{y}_{N};0)$ is symmetric under the permutations of $\vec{y}_{N}$.
Inserting the initial condition (\ref{iniGeneral}) into Eq.\ (\ref{PNXT}) gives
\begin{equation}
\notag
f^{(N)}(\vec{x}_{N};t) = N!\!\! \int\limits_{\mathcal{R}_{N}} \!\! {\rm d}\vec{y}_{N}\,
p^{(N)}(\vec{x}_{N};t\,|\,\vec{y}_{N};0)\prod_{n=1}^{N} f(y_{n})\,\,.
\end{equation}
We now insert $p^{(N)}(\vec{x}_{N};t\,|\,\vec{y}_{N};0)$ as given by  Eq.\ (\ref{permfinal}), interchange the summation and integration, and use the symmetry of $p^{(N)}(\vec{x}_{N};t\,|\,\vec{y}_{N};0)$ (cf. Eq.\ (\ref{integraceG})). We obtain
\begin{eqnarray}
\notag
f^{(N)}(\vec{x}_{N};t) &=& \sum_{m=1}^{N!}
\int_{0}^{+\infty} \!\!\!\!\!\!\!\!\!{\rm d}y_{N}\,\, p(x_{\pi_{m}(N)};t\,|\,y_{N};0) f(y_{N}) \\
\notag
 && \times \int_{0}^{+\infty} \!\!\!\!\!\!\!\!\!{\rm d}y_{N-1}\,\, p(x_{\pi_{m}(N-1)};t\,|\,y_{N-1};0) f(y_{N-1})\, ... \\
\notag
&& \times \int_{0}^{+\infty} \!\!\!\!\!\!\!\!\! {\rm d}y_{1}\,\, p(x_{\pi_{m}(1)};t\,|\,y_{1};0) f(y_{1})\\
&=& \sum_{m=1}^{N!} p(x_{\pi_{m}(N)}; t)p(x_{\pi_{m}(N-1)}; t) ... p(x_{\pi_{m}(1)}; t)\,\,,
\notag
\end{eqnarray}
where the last equality follows from the definition (\ref{pDEF}).
The last sum comprises $N!$ identical summands and hence we arrive at Eq.\ (\ref{PNXTproduct}).
\section{Derivation of $S_{n}(t)$}
\label{app:SPC}

Let us define the auxiliary function $B(x,t)=S(t)-F(x,t)$ with the properties: $B(0,t)=S(t)$, $B(+\infty,t)=0$, $p(x;t)=-\frac{\partial}{\partial x}B(x,t)$. Then we can rewrite Eq.\ (\ref{Sndefin}) as
\begin{equation}
S_{n}(t) = - n \binom{N}{n}\int_{0}^{+\infty}\!\!\!\!\!\! {\rm d} x\,\, \frac{\partial B}{\partial x}\, B^{n-1}(1-B)^{N-n}\,\,.
\end{equation}
Using the binomial theorem we expand $(1-B)^{N-n}$ and we insert the unity in the form $(n+k)/(n+k)$. Then we get
\begin{equation}
S_{n}(t) = - n \binom{N}{n}\sum_{k=0}^{N-n} \frac{\left(-1\right)^{k}}{n+k} \binom{N-n}{k}
\int_{0}^{+\infty}\!\!\!\!\!\! {\rm d} x\,\, \frac{\partial B^{n+k}}{\partial x}\, \,\,,
\label{C2}
\end{equation}
where the definite integral equals $\left(S(t)\right)^{n+k}$. On the whole, we have proven Eq.\ (\ref{survn}) from the main text. Moreover, the sum in Eq.\ (\ref{survn}) can be represented by the definite integral\cite{Prudnikov}
\begin{equation}
S_{n}(t) =  n \binom{N}{n} \int_{0}^{S(t)}\!\!\!\!\!\! {\rm d} z\,\, z^{n-1} (1-z)^{N-n}\, \,\,.
\label{C3}
\end{equation}
The integration by parts directly yields Eq.\ (\ref{rozdilS}).

%

\begin{thebibliography}{36}%
\makeatletter
\providecommand \@ifxundefined [1]{%
 \@ifx{#1\undefined}
}%
\providecommand \@ifnum [1]{%
 \ifnum #1\expandafter \@firstoftwo
 \else \expandafter \@secondoftwo
 \fi
}%
\providecommand \@ifx [1]{%
 \ifx #1\expandafter \@firstoftwo
 \else \expandafter \@secondoftwo
 \fi
}%
\providecommand \natexlab [1]{#1}%
\providecommand \enquote  [1]{``#1''}%
\providecommand \bibnamefont  [1]{#1}%
\providecommand \bibfnamefont [1]{#1}%
\providecommand \citenamefont [1]{#1}%
\providecommand \href@noop [0]{\@secondoftwo}%
\providecommand \href [0]{\begingroup \@sanitize@url \@href}%
\providecommand \@href[1]{\@@startlink{#1}\@@href}%
\providecommand \@@href[1]{\endgroup#1\@@endlink}%
\providecommand \@sanitize@url [0]{\catcode `\\12\catcode `\$12\catcode
  `\&12\catcode `\#12\catcode `\^12\catcode `\_12\catcode `\%12\relax}%
\providecommand \@@startlink[1]{}%
\providecommand \@@endlink[0]{}%
\providecommand \url  [0]{\begingroup\@sanitize@url \@url }%
\providecommand \@url [1]{\endgroup\@href {#1}{\urlprefix }}%
\providecommand \urlprefix  [0]{URL }%
\providecommand \Eprint [0]{\href }%
\providecommand \doibase [0]{http://dx.doi.org/}%
\providecommand \selectlanguage [0]{\@gobble}%
\providecommand \bibinfo  [0]{\@secondoftwo}%
\providecommand \bibfield  [0]{\@secondoftwo}%
\providecommand \translation [1]{[#1]}%
\providecommand \BibitemOpen [0]{}%
\providecommand \bibitemStop [0]{}%
\providecommand \bibitemNoStop [0]{.\EOS\space}%
\providecommand \EOS [0]{\spacefactor3000\relax}%
\providecommand \BibitemShut  [1]{\csname bibitem#1\endcsname}%
\let\auto@bib@innerbib\@empty
\bibitem [{\citenamefont {Hodgkin}\ and\ \citenamefont
  {Keynes}(1955)}]{Hodgkin}%
  \BibitemOpen
  \bibfield  {author} {\bibinfo {author} {\bibfnamefont {A.~L.}\ \bibnamefont
  {Hodgkin}}\ and\ \bibinfo {author} {\bibfnamefont {R.~D.}\ \bibnamefont
  {Keynes}},\ }\href@noop {} {\bibfield  {journal} {\bibinfo  {journal} {J.
  Physiol.}\ }\textbf {\bibinfo {volume} {128}},\ \bibinfo {pages} {61}
  (\bibinfo {year} {1955})}\BibitemShut {NoStop}%
\bibitem [{\citenamefont {Hahn}, \citenamefont {K{\" a}rger},\ and\
  \citenamefont {Kukla}(1996)}]{HahnKarger}%
  \BibitemOpen
  \bibfield  {author} {\bibinfo {author} {\bibfnamefont {K.}~\bibnamefont
  {Hahn}}, \bibinfo {author} {\bibfnamefont {J.}~\bibnamefont {K{\" a}rger}}, \
  and\ \bibinfo {author} {\bibfnamefont {V.}~\bibnamefont {Kukla}},\
  }\href@noop {} {\bibfield  {journal} {\bibinfo  {journal} {Phys. Rev. Lett.}\
  }\textbf {\bibinfo {volume} {76}},\ \bibinfo {pages} {2762} (\bibinfo {year}
  {1996})}\BibitemShut {NoStop}%
\bibitem [{\citenamefont {K{\" a}rger}\ and\ \citenamefont
  {Ruthven}(1992)}]{KargerBOOK}%
  \BibitemOpen
  \bibfield  {author} {\bibinfo {author} {\bibfnamefont {J.}~\bibnamefont {K{\"
  a}rger}}\ and\ \bibinfo {author} {\bibfnamefont {M.}~\bibnamefont
  {Ruthven}},\ }\href@noop {} {\emph {\bibinfo {title} {Diffusion in Zeolites
  and in Other Microporous Solids}}}\ (\bibinfo  {publisher} {Wiley},\ \bibinfo
  {address} {New York},\ \bibinfo {year} {1992})\BibitemShut {NoStop}%
\bibitem [{\citenamefont {Li}, \citenamefont {Berg},\ and\ \citenamefont
  {Elf}(2009)}]{DNAnature}%
  \BibitemOpen
  \bibfield  {author} {\bibinfo {author} {\bibfnamefont {G.-W.}\ \bibnamefont
  {Li}}, \bibinfo {author} {\bibfnamefont {O.~G.}\ \bibnamefont {Berg}}, \ and\
  \bibinfo {author} {\bibfnamefont {J.}~\bibnamefont {Elf}},\ }\href@noop {}
  {\bibfield  {journal} {\bibinfo  {journal} {Nature Physics}\ }\textbf
  {\bibinfo {volume} {5}},\ \bibinfo {pages} {294} (\bibinfo {year}
  {2009})}\BibitemShut {NoStop}%
\bibitem [{\citenamefont {Berg}, \citenamefont {Winter},\ and\ \citenamefont
  {Hippel}(1981)}]{Hippel}%
  \BibitemOpen
  \bibfield  {author} {\bibinfo {author} {\bibfnamefont {O.~G.}\ \bibnamefont
  {Berg}}, \bibinfo {author} {\bibfnamefont {R.~B.}\ \bibnamefont {Winter}}, \
  and\ \bibinfo {author} {\bibfnamefont {P.~H.~V.}\ \bibnamefont {Hippel}},\
  }\href@noop {} {\bibfield  {journal} {\bibinfo  {journal} {Biochemistry}\
  }\textbf {\bibinfo {volume} {20}},\ \bibinfo {pages} {6929} (\bibinfo {year}
  {1981})}\BibitemShut {NoStop}%
\bibitem [{\citenamefont {Halpin-Healy}\ and\ \citenamefont
  {Zhang}(1995)}]{Halpin}%
  \BibitemOpen
  \bibfield  {author} {\bibinfo {author} {\bibfnamefont {T.}~\bibnamefont
  {Halpin-Healy}}\ and\ \bibinfo {author} {\bibfnamefont {Y.-C.}\ \bibnamefont
  {Zhang}},\ }\href@noop {} {\bibfield  {journal} {\bibinfo  {journal} {Phys.
  Rep.}\ }\textbf {\bibinfo {volume} {254}},\ \bibinfo {pages} {215} (\bibinfo
  {year} {1995})}\BibitemShut {NoStop}%
\bibitem [{\citenamefont {Mukherjee}\ \emph {et~al.}(2007)\citenamefont
  {Mukherjee}, \citenamefont {Mait}, \citenamefont {Dasgupta},\ and\
  \citenamefont {Sood}}]{nanoWaterI}%
  \BibitemOpen
  \bibfield  {author} {\bibinfo {author} {\bibfnamefont {B.}~\bibnamefont
  {Mukherjee}}, \bibinfo {author} {\bibfnamefont {P.~K.}\ \bibnamefont {Mait}},
  \bibinfo {author} {\bibfnamefont {C.}~\bibnamefont {Dasgupta}}, \ and\
  \bibinfo {author} {\bibfnamefont {A.~K.}\ \bibnamefont {Sood}},\ }\href@noop
  {} {\bibfield  {journal} {\bibinfo  {journal} {Nanosci. Nanotechnol.}\
  }\textbf {\bibinfo {volume} {7}},\ \bibinfo {pages} {1796} (\bibinfo {year}
  {2007})}\BibitemShut {NoStop}%
\bibitem [{\citenamefont {Cheng}\ and\ \citenamefont
  {Bowers}(2007)}]{nanoWaterII}%
  \BibitemOpen
  \bibfield  {author} {\bibinfo {author} {\bibfnamefont {C.-Y.}\ \bibnamefont
  {Cheng}}\ and\ \bibinfo {author} {\bibfnamefont {C.~R.}\ \bibnamefont
  {Bowers}},\ }\href@noop {} {\bibfield  {journal} {\bibinfo  {journal}
  {ChemPhysChem}\ }\textbf {\bibinfo {volume} {8}},\ \bibinfo {pages} {2077}
  (\bibinfo {year} {2007})}\BibitemShut {NoStop}%
\bibitem [{\citenamefont {Kharkyanen}, \citenamefont {Yesylevskyy},\ and\
  \citenamefont {Berezetskaya}(2010)}]{Superionic}%
  \BibitemOpen
  \bibfield  {author} {\bibinfo {author} {\bibfnamefont {V.~N.}\ \bibnamefont
  {Kharkyanen}}, \bibinfo {author} {\bibfnamefont {S.~O.}\ \bibnamefont
  {Yesylevskyy}}, \ and\ \bibinfo {author} {\bibfnamefont {N.~M.}\ \bibnamefont
  {Berezetskaya}},\ }\href@noop {} {\bibfield  {journal} {\bibinfo  {journal}
  {Phys. Rev. E}\ }\textbf {\bibinfo {volume} {82}},\ \bibinfo {pages} {051103}
  (\bibinfo {year} {2010})}\BibitemShut {NoStop}%
\bibitem [{\citenamefont {Richards}(1977)}]{SuperionicII}%
  \BibitemOpen
  \bibfield  {author} {\bibinfo {author} {\bibfnamefont {P.~M.}\ \bibnamefont
  {Richards}},\ }\href@noop {} {\bibfield  {journal} {\bibinfo  {journal}
  {Phys. Rev. B}\ }\textbf {\bibinfo {volume} {16}},\ \bibinfo {pages} {1393}
  (\bibinfo {year} {1977})}\BibitemShut {NoStop}%
\bibitem [{\citenamefont {Dekker}(2007)}]{SSnanopores}%
  \BibitemOpen
  \bibfield  {author} {\bibinfo {author} {\bibfnamefont {C.}~\bibnamefont
  {Dekker}},\ }\href@noop {} {\bibfield  {journal} {\bibinfo  {journal} {Nature
  Nanotechnol.}\ }\textbf {\bibinfo {volume} {2}},\ \bibinfo {pages} {209}
  (\bibinfo {year} {2007})}\BibitemShut {NoStop}%
\bibitem [{\citenamefont {Lutz}, \citenamefont {Kollmann},\ and\ \citenamefont
  {Bechinger}(2004)}]{BechCircle}%
  \BibitemOpen
  \bibfield  {author} {\bibinfo {author} {\bibfnamefont {C.}~\bibnamefont
  {Lutz}}, \bibinfo {author} {\bibfnamefont {M.}~\bibnamefont {Kollmann}}, \
  and\ \bibinfo {author} {\bibfnamefont {C.}~\bibnamefont {Bechinger}},\
  }\href@noop {} {\bibfield  {journal} {\bibinfo  {journal} {Phys. Rev. Lett.}\
  }\textbf {\bibinfo {volume} {93}},\ \bibinfo {pages} {026001} (\bibinfo
  {year} {2004})}\BibitemShut {NoStop}%
\bibitem [{\citenamefont {Lutz}, \citenamefont {Kollmann},\ and\ \citenamefont
  {Bechinger}(2000)}]{BechPhoto}%
  \BibitemOpen
  \bibfield  {author} {\bibinfo {author} {\bibfnamefont {C.}~\bibnamefont
  {Lutz}}, \bibinfo {author} {\bibfnamefont {M.}~\bibnamefont {Kollmann}}, \
  and\ \bibinfo {author} {\bibfnamefont {C.}~\bibnamefont {Bechinger}},\
  }\href@noop {} {\bibfield  {journal} {\bibinfo  {journal} {Science}\ }\textbf
  {\bibinfo {volume} {287}},\ \bibinfo {pages} {625} (\bibinfo {year}
  {2000})}\BibitemShut {NoStop}%
\bibitem [{\citenamefont {Chou}, \citenamefont {Eng},\ and\ \citenamefont
  {Robert}(2006)}]{colloidpolym}%
  \BibitemOpen
  \bibfield  {author} {\bibinfo {author} {\bibfnamefont {C.-Y.}\ \bibnamefont
  {Chou}}, \bibinfo {author} {\bibfnamefont {B.~C.}\ \bibnamefont {Eng}}, \
  and\ \bibinfo {author} {\bibfnamefont {M.}~\bibnamefont {Robert}},\
  }\href@noop {} {\bibfield  {journal} {\bibinfo  {journal} {J. Chem. Phys.}\
  }\textbf {\bibinfo {volume} {124}},\ \bibinfo {pages} {044902} (\bibinfo
  {year} {2006})}\BibitemShut {NoStop}%
\bibitem [{\citenamefont {Coupier}, \citenamefont {Jean},\ and\ \citenamefont
  {Guthmann}(2006)}]{ChargedSpheres}%
  \BibitemOpen
  \bibfield  {author} {\bibinfo {author} {\bibfnamefont {G.}~\bibnamefont
  {Coupier}}, \bibinfo {author} {\bibfnamefont {M.~S.}\ \bibnamefont {Jean}}, \
  and\ \bibinfo {author} {\bibfnamefont {C.}~\bibnamefont {Guthmann}},\
  }\href@noop {} {\bibfield  {journal} {\bibinfo  {journal} {Phys. Rev. E}\
  }\textbf {\bibinfo {volume} {73}},\ \bibinfo {pages} {031112} (\bibinfo
  {year} {2006})}\BibitemShut {NoStop}%
\bibitem [{\citenamefont {Delfau}\ \emph {et~al.}(2010)\citenamefont {Delfau},
  \citenamefont {Coste}, \citenamefont {Even},\ and\ \citenamefont
  {Jean}}]{BallsOFsteel}%
  \BibitemOpen
  \bibfield  {author} {\bibinfo {author} {\bibfnamefont {J.~B.}\ \bibnamefont
  {Delfau}}, \bibinfo {author} {\bibfnamefont {C.}~\bibnamefont {Coste}},
  \bibinfo {author} {\bibfnamefont {C.}~\bibnamefont {Even}}, \ and\ \bibinfo
  {author} {\bibfnamefont {M.~S.}\ \bibnamefont {Jean}},\ }\href@noop {}
  {\bibfield  {journal} {\bibinfo  {journal} {Phys. Rev. E}\ }\textbf {\bibinfo
  {volume} {82}},\ \bibinfo {pages} {031201} (\bibinfo {year}
  {2010})}\BibitemShut {NoStop}%
\bibitem [{\citenamefont {Harris}(1965)}]{Harris}%
  \BibitemOpen
  \bibfield  {author} {\bibinfo {author} {\bibfnamefont {T.~E.}\ \bibnamefont
  {Harris}},\ }\href@noop {} {\bibfield  {journal} {\bibinfo  {journal} {J.
  Appl. Prob.}\ }\textbf {\bibinfo {volume} {2}},\ \bibinfo {pages} {323}
  (\bibinfo {year} {1965})}\BibitemShut {NoStop}%
\bibitem [{\citenamefont {Barkai}\ and\ \citenamefont
  {Silbey}(2010)}]{BarSil2010}%
  \BibitemOpen
  \bibfield  {author} {\bibinfo {author} {\bibfnamefont {E.}~\bibnamefont
  {Barkai}}\ and\ \bibinfo {author} {\bibfnamefont {R.}~\bibnamefont
  {Silbey}},\ }\href@noop {} {\bibfield  {journal} {\bibinfo  {journal} {Phys.
  Rev. E}\ }\textbf {\bibinfo {volume} {81}},\ \bibinfo {pages} {041129}
  (\bibinfo {year} {2010})}\BibitemShut {NoStop}%
\bibitem [{\citenamefont {R{\" o}denbeck}, \citenamefont {K{\" a}rger},\ and\
  \citenamefont {Hahn}(1998)}]{KargerHahn}%
  \BibitemOpen
  \bibfield  {author} {\bibinfo {author} {\bibfnamefont {C.}~\bibnamefont {R{\"
  o}denbeck}}, \bibinfo {author} {\bibfnamefont {J.}~\bibnamefont {K{\"
  a}rger}}, \ and\ \bibinfo {author} {\bibfnamefont {K.}~\bibnamefont {Hahn}},\
  }\href@noop {} {\bibfield  {journal} {\bibinfo  {journal} {Phys. Rev. E}\
  }\textbf {\bibinfo {volume} {57}},\ \bibinfo {pages} {4382} (\bibinfo {year}
  {1998})}\BibitemShut {NoStop}%
\bibitem [{\citenamefont {Barkai}\ and\ \citenamefont {Silbey}(2008)}]{Kumar}%
  \BibitemOpen
  \bibfield  {author} {\bibinfo {author} {\bibfnamefont {E.}~\bibnamefont
  {Barkai}}\ and\ \bibinfo {author} {\bibfnamefont {R.}~\bibnamefont
  {Silbey}},\ }\href@noop {} {\bibfield  {journal} {\bibinfo  {journal} {Phys.
  Rev. E}\ }\textbf {\bibinfo {volume} {78}},\ \bibinfo {pages} {021133}
  (\bibinfo {year} {2008})}\BibitemShut {NoStop}%
\bibitem [{\citenamefont {Lizana}\ and\ \citenamefont {Ambj{\"
  o}rnsson}(2009)}]{SFDLizana}%
  \BibitemOpen
  \bibfield  {author} {\bibinfo {author} {\bibfnamefont {L.}~\bibnamefont
  {Lizana}}\ and\ \bibinfo {author} {\bibfnamefont {T.}~\bibnamefont {Ambj{\"
  o}rnsson}},\ }\href@noop {} {\bibfield  {journal} {\bibinfo  {journal} {Phys.
  Rev. E}\ }\textbf {\bibinfo {volume} {80}},\ \bibinfo {pages} {051103}
  (\bibinfo {year} {2009})}\BibitemShut {NoStop}%
\bibitem [{\citenamefont {Ryabov}\ and\ \citenamefont
  {Chvosta}(2011)}]{RC2011}%
  \BibitemOpen
  \bibfield  {author} {\bibinfo {author} {\bibfnamefont {A.}~\bibnamefont
  {Ryabov}}\ and\ \bibinfo {author} {\bibfnamefont {P.}~\bibnamefont
  {Chvosta}},\ }\href@noop {} {\bibfield  {journal} {\bibinfo  {journal} {Phys.
  Rev. E}\ }\textbf {\bibinfo {volume} {83}},\ \bibinfo {pages} {020106}
  (\bibinfo {year} {2011})}\BibitemShut {NoStop}%
\bibitem [{\citenamefont {Liu}\ \emph {et~al.}(2011)\citenamefont {Liu},
  \citenamefont {Wang}, \citenamefont {Ackerman}, \citenamefont {Slowing},
  \citenamefont {Pruski}, \citenamefont {Chen}, \citenamefont {Lin},\ and\
  \citenamefont {Evans}}]{DaJiangLiu}%
  \BibitemOpen
  \bibfield  {author} {\bibinfo {author} {\bibfnamefont {D.-J.}\ \bibnamefont
  {Liu}}, \bibinfo {author} {\bibfnamefont {J.}~\bibnamefont {Wang}}, \bibinfo
  {author} {\bibfnamefont {D.~M.}\ \bibnamefont {Ackerman}}, \bibinfo {author}
  {\bibfnamefont {I.~I.}\ \bibnamefont {Slowing}}, \bibinfo {author}
  {\bibfnamefont {M.}~\bibnamefont {Pruski}}, \bibinfo {author} {\bibfnamefont
  {H.-T.}\ \bibnamefont {Chen}}, \bibinfo {author} {\bibfnamefont {V.~S.-Y.}\
  \bibnamefont {Lin}}, \ and\ \bibinfo {author} {\bibfnamefont {J.~W.}\
  \bibnamefont {Evans}},\ }\href@noop {} {\bibfield  {journal} {\bibinfo
  {journal} {ACS Catal.}\ }\textbf {\bibinfo {volume} {1}},\ \bibinfo {pages}
  {751} (\bibinfo {year} {2011})}\BibitemShut {NoStop}%
\bibitem [{\citenamefont {Dorsaz}\ \emph {et~al.}(2010)\citenamefont {Dorsaz},
  \citenamefont {Michele}, \citenamefont {Piazza}, \citenamefont {Rios},\ and\
  \citenamefont {Foffi}}]{Smoluch}%
  \BibitemOpen
  \bibfield  {author} {\bibinfo {author} {\bibfnamefont {N.}~\bibnamefont
  {Dorsaz}}, \bibinfo {author} {\bibfnamefont {C.~D.}\ \bibnamefont {Michele}},
  \bibinfo {author} {\bibfnamefont {F.}~\bibnamefont {Piazza}}, \bibinfo
  {author} {\bibfnamefont {P.~D.~L.}\ \bibnamefont {Rios}}, \ and\ \bibinfo
  {author} {\bibfnamefont {G.}~\bibnamefont {Foffi}},\ }\href@noop {}
  {\bibfield  {journal} {\bibinfo  {journal} {Phys. Rev. Lett.}\ }\textbf
  {\bibinfo {volume} {105}},\ \bibinfo {pages} {120601} (\bibinfo {year}
  {2010})}\BibitemShut {NoStop}%
\bibitem [{\citenamefont {Park}, \citenamefont {Kim},\ and\ \citenamefont
  {Shin}(2003)}]{JoonhoPark}%
  \BibitemOpen
  \bibfield  {author} {\bibinfo {author} {\bibfnamefont {J.}~\bibnamefont
  {Park}}, \bibinfo {author} {\bibfnamefont {H.}~\bibnamefont {Kim}}, \ and\
  \bibinfo {author} {\bibfnamefont {K.~J.}\ \bibnamefont {Shin}},\ }\href@noop
  {} {\bibfield  {journal} {\bibinfo  {journal} {J. Chem. Phys.}\ }\textbf
  {\bibinfo {volume} {118}},\ \bibinfo {pages} {9697} (\bibinfo {year}
  {2003})}\BibitemShut {NoStop}%
\bibitem [{\citenamefont {Seki}, \citenamefont {Wojcik},\ and\ \citenamefont
  {Tachiya}(2011)}]{Kazuhiko}%
  \BibitemOpen
  \bibfield  {author} {\bibinfo {author} {\bibfnamefont {K.}~\bibnamefont
  {Seki}}, \bibinfo {author} {\bibfnamefont {M.}~\bibnamefont {Wojcik}}, \ and\
  \bibinfo {author} {\bibfnamefont {M.}~\bibnamefont {Tachiya}},\ }\href@noop
  {} {\bibfield  {journal} {\bibinfo  {journal} {J. Chem. Phys.}\ }\textbf
  {\bibinfo {volume} {134}},\ \bibinfo {pages} {094506} (\bibinfo {year}
  {2011})}\BibitemShut {NoStop}%
\bibitem [{\citenamefont {Redner}(2001)}]{Redner}%
  \BibitemOpen
  \bibfield  {author} {\bibinfo {author} {\bibfnamefont {S.}~\bibnamefont
  {Redner}},\ }\href@noop {} {\emph {\bibinfo {title} {A Guide to First-Passage
  Processes}}}\ (\bibinfo  {publisher} {Cambridge University Press},\ \bibinfo
  {year} {2001})\BibitemShut {NoStop}%
\bibitem [{\citenamefont {Abramowitz}\ and\ \citenamefont
  {Stegun}(1972)}]{AbrStegun}%
  \BibitemOpen
  \bibinfo {editor} {\bibfnamefont {M.}~\bibnamefont {Abramowitz}}\ and\
  \bibinfo {editor} {\bibfnamefont {I.~A.}\ \bibnamefont {Stegun}},\ eds.,\
  \enquote {\bibinfo {title} {Handbook of mathematical functions with formulas,
  graphs, and mathematical tables},}\ \ (\bibinfo  {publisher} {Dover},\
  \bibinfo {address} {New York},\ \bibinfo {year} {1972})\ Chap.~\bibinfo
  {chapter} {7}, p.\ \bibinfo {pages} {297}\BibitemShut {NoStop}%
\bibitem [{\citenamefont {Duffy}(2004)}]{Duffy}%
  \BibitemOpen
  \bibfield  {author} {\bibinfo {author} {\bibfnamefont {D.~G.}\ \bibnamefont
  {Duffy}},\ }\href@noop {} {\emph {\bibinfo {title} {Transform Methods for
  Solving Partial Differential Equations, Second Edition}}}\ (\bibinfo
  {publisher} {Chapman and Hall/CRC},\ \bibinfo {year} {2004})\BibitemShut
  {NoStop}%
\bibitem [{\citenamefont {Karlin}\ and\ \citenamefont {Taylor}(1981)}]{karlin}%
  \BibitemOpen
  \bibfield  {author} {\bibinfo {author} {\bibfnamefont {S.}~\bibnamefont
  {Karlin}}\ and\ \bibinfo {author} {\bibfnamefont {H.~M.}\ \bibnamefont
  {Taylor}},\ }\href@noop {} {\emph {\bibinfo {title} {A Second Couse in
  Stochastic Processes}}}\ (\bibinfo  {publisher} {Accademic Press},\ \bibinfo
  {address} {San Diego},\ \bibinfo {year} {1981})\BibitemShut {NoStop}%
\bibitem [{\citenamefont {Minc}(1978)}]{permanents}%
  \BibitemOpen
  \bibfield  {author} {\bibinfo {author} {\bibfnamefont {H.}~\bibnamefont
  {Minc}},\ }\href@noop {} {\emph {\bibinfo {title} {Permanents}}},\ \bibinfo
  {series} {Encyclopedia of mathematics and its applications}, Vol.~\bibinfo
  {volume} {6}\ (\bibinfo  {publisher} {Addison-Wesley},\ \bibinfo {address}
  {Reading, Massachusetts},\ \bibinfo {year} {1978})\BibitemShut {NoStop}%
\bibitem [{Note1()}]{Note1}%
  \BibitemOpen
  \bibinfo {note} {$f(t) \sim g(t), \protect \tmspace +\thinmuskip
  {.1667em}\protect \tmspace +\thinmuskip {.1667em} \protect \mathrm {as}
  \protect \tmspace +\thinmuskip {.1667em}\protect \tmspace +\thinmuskip
  {.1667em} t \rightarrow \infty \protect \tmspace +\thinmuskip {.1667em}$
  means $ \displaystyle \protect \qopname \relax m{lim}_{t \to \infty }
  f(t)/g(t) = 1$.}\BibitemShut {Stop}%
\bibitem [{\citenamefont {Weiss}, \citenamefont {Shuler},\ and\ \citenamefont
  {Lindenberg}(1983)}]{Lindenberg}%
  \BibitemOpen
  \bibfield  {author} {\bibinfo {author} {\bibfnamefont {G.~H.}\ \bibnamefont
  {Weiss}}, \bibinfo {author} {\bibfnamefont {K.~E.}\ \bibnamefont {Shuler}}, \
  and\ \bibinfo {author} {\bibfnamefont {K.}~\bibnamefont {Lindenberg}},\
  }\href@noop {} {\bibfield  {journal} {\bibinfo  {journal} {J. Stat. Phys.}\
  }\textbf {\bibinfo {volume} {31}},\ \bibinfo {pages} {255} (\bibinfo {year}
  {1983})}\BibitemShut {NoStop}%
\bibitem [{\citenamefont {Yuste}\ and\ \citenamefont
  {Lindenberg}(1996)}]{Yuste}%
  \BibitemOpen
  \bibfield  {author} {\bibinfo {author} {\bibfnamefont {S.~B.}\ \bibnamefont
  {Yuste}}\ and\ \bibinfo {author} {\bibfnamefont {K.}~\bibnamefont
  {Lindenberg}},\ }\href@noop {} {\bibfield  {journal} {\bibinfo  {journal} {J.
  Stat. Phys.}\ }\textbf {\bibinfo {volume} {85}},\ \bibinfo {pages} {501}
  (\bibinfo {year} {1996})}\BibitemShut {NoStop}%
\bibitem [{\citenamefont {Ben-Naim}\ and\ \citenamefont
  {Krapivsky}(2010)}]{Krapivsky}%
  \BibitemOpen
  \bibfield  {author} {\bibinfo {author} {\bibfnamefont {E.}~\bibnamefont
  {Ben-Naim}}\ and\ \bibinfo {author} {\bibfnamefont {P.~L.}\ \bibnamefont
  {Krapivsky}},\ }\href@noop {} {\bibfield  {journal} {\bibinfo  {journal} {J.
  Phys. A: Math. Theor.}\ }\textbf {\bibinfo {volume} {43}},\ \bibinfo {pages}
  {495008} (\bibinfo {year} {2010})}\BibitemShut {NoStop}%
\bibitem [{\citenamefont {Prudnikov}, \citenamefont {Brychkov},\ and\
  \citenamefont {Marichev}(1986)}]{Prudnikov}%
  \BibitemOpen
  \bibfield  {author} {\bibinfo {author} {\bibfnamefont {A.~P.}\ \bibnamefont
  {Prudnikov}}, \bibinfo {author} {\bibfnamefont {Y.~A.}\ \bibnamefont
  {Brychkov}}, \ and\ \bibinfo {author} {\bibfnamefont {O.~I.}\ \bibnamefont
  {Marichev}},\ }\href@noop {} {\emph {\bibinfo {title} {Integrals and Series,
  Vol. 1: Elementary Functions}}}\ (\bibinfo  {publisher} {Gordon and Breach},\
  \bibinfo {year} {1986})\BibitemShut {NoStop}%
\end{thebibliography}
\end{document}